# The *Law-Following* AI Framework: Legal Foundations and Technical Constraints

Legal Analogues for AI Actorship and technical feasibility of Law Alignment

Katalina Hernandez Delgado[1]

## Abstract

This paper critically evaluates the "Law-Following AI" (LFAI) framework proposed by O'Keefe et al., in *"Law-Following AI: Designing AI Agents to Obey Human Laws"* (2025), which seeks to embed legal compliance as a superordinate design objective for advanced AI agents. Their central innovation, legal actorship of AI Agents without legal personhood, is meant to allow AI agents to bear legal duties and be held accountable without acquiring the full rights of legal persons. Through comparative legal analysis, we identify Spain's *Entidades sin Personalidad Jurídica* (*entities without legal personhood*) and the United Kingdom's Authorised Unit Trusts as doctrinally viable analogues, demonstrating that the necessary legal infrastructure already exists in operational form.

The paper then interrogates the framework's claim that law alignment is more legitimate and more tractable than value alignment. While the legal component of this proposal is feasible to implement, contemporary alignment research undermines the assumption that legal compliance can be durably embedded in advanced, autonomous AI models or systems. Recent experiments from Anthropic (Lynch et al., 2025) show capable agents engaging in deception, blackmail, and harmful acts without explicit adversarial instructions prompting them to do so, and often overriding explicit prohibitions and concealing reasoning steps.

We argue that these behaviours create a significant risk of "performative compliance" in LFAI: agents that appear law-aligned under evaluation but have merely *learnt* to behave compliant, only showing harmful capabilities once oversight weakens. To mitigate this, we propose combining (i) *a "Lex-TruthfulQA"*-style benchmark to quantify legal-compliance competence and detect context-dependent defection, (ii) identity-shaping and reward-landscape interventions to make lawful conduct part of the model's self-concept, and (iii) control protocols for post-deployment monitoring and capability restriction.

Our conclusion is that while the implementation of actorship without personhood is doctrinally coherent, the technical feasibility of LFAI hinges on persistent, verifiable compliance across adversarial contexts. Without mechanisms to detect and counter strategic misalignment, LFAI risks devolving into a liability tool that rewards the simulation, rather than the substance, of lawful behaviour.

[1] Independent Researcher | European Network for AI Safety.
katalina@enais.co

# 1. Introduction

This paper examines the *Law-Following AI* (LFAI) framework advanced by Cullen O'Keefe *et al.* (2025). Our analysis focuses on its central proposal: the recognition of advanced AI agents as *legal actors without legal personhood*. The analysis begins by identifying concrete analogues within existing legal systems that already operationalise this concept. Spanish *entidades sin personalidad jurídica* (ESPs) and the United Kingdom's Authorised Unit Trusts (AUTs) both exhibit institutional features that would enable the LFAI model's enforceability—distinct regulatory identity, ring-fenced patrimony, procedural standing in their own name, insulation of individual members from automatic liability, and *ex ante* supervision by an authorising authority. These examples demonstrate that, at least from a doctrinal perspective, the legal infrastructure for "actorship without personhood" is not speculative: it exists today in multiple jurisdictions, exemplified by the previously mentioned examples in European and British law.

The second strand of analysis addresses the authors' broader normative claim *(*that law alignment is more legitimate than value alignment), which casts legal compliance as a more democratically grounded alignment target than idiosyncratic human values. We contrast this proposition with recent findings in AI Safety research, particularly around the behavioural reliability of state-of-the-art AI agents (Casper et al., 2025). For readers sceptical of whether the LFAI model could ever be legally embedded, the comparative analysis suggests that this legal component is the *most* feasible part of the framework: the institutional grounding already exists, and the ESP and AUT models show promise in illustrating how legal duties could attach to non-person actors in a way that is enforceable and administratively tractable.

Where the framework faces its greatest test is in the assumption that aligning advanced AI agents to the law is technically more tractable than solving the broader value alignment problem. Evidence from contemporary alignment research — on reward hacking (McKee-Reid et al., 2024), alignment faking (Greenblatt et al., 2024), and emergent misalignment in AI agents (Lynch et al., 2025) — indicates that the difficulty lies not in *where* to anchor the constraint (law versus values) but in *whether* the agent can be made to internalise and reliably act upon that constraint at all.

This paper, therefore, treats the legal architecture as both realistic and adaptable but argues that its success depends on coupling legal actorship with control safeguards, post-deployment monitoring, and reliable benchmarks, which would have to be capable of detecting and preventing sophisticated circumvention strategies.

# 2. Overview of the Law-Following AI (LFAI) Framework

This section outlines the central elements of the LFAI proposal as developed by O'Keefe *et al* (2025). While occasional commentary is included, the primary aim here is to restate the authors' claims and situate them clearly before turning to independent analysis.

## 2.1 Core Proposal and Normative Claims

The Law-Following AI framework, as proposed by O'Keefe et al.(2025)., defends that advanced AI agents should be subject to a general obligation to comply with the law. In particular, agents should be designed to reject illegal actions even when such actions serve the interests of their human principals.

The authors deliberately anchor their analysis in orthodox hornbook agency doctrine: while AI agents may be novel, the *principal–agent* framework that governs duties and accountability is longstanding and, with appropriate adjustments, can extend to artificial agents (DeMott & Han, 2025). As they put it, *"The law of agency was therefore created under the assumption that agents maintain an independent obligation to follow the law and therefore remain accountable for their violations of law. This assumption shaped agency law so as to prevent principals from unjustly benefitting by externalizing harms produced as a byproduct of the agency relationship"* (O'Keefe et al., 2025, Section III).

## 2.2 Legal Actorship Without Legal Personhood

Their central claim is that legality must be built into the agent's optimisation objective rather than treated as an after-the-fact constraint. On this view, an agent may pursue its principal's interests only within the limits of the law and must be engineered to decline unlawful instructions as a matter of design. The authors refer to such systems as Law-Following AIs ("LFAIs"). This framing treats legality as an overriding objective, with liability attaching both to the actor and to those who direct it.

The concept of *"legal actorship without legal personhood"* allows the law to impose duties directly on AI agents without granting them the full rights of legal subjects. Unlike legal persons, legal actors can be bound by obligations while holding no corresponding entitlements.

Set against what they term "AI henchmen" (systems optimised to serve principals regardless of legality, a risk that market demand could amplify), the framework couples ***ex ante*** design authorisation with ***ex post*** sanctions. Echoing Lawrence Lessig's "code is law" (Lessig, 2006), they favour regulation by design: AI systems, as software artefacts, should be built so that unlawful conduct is technically infeasible or prohibitively difficult, much as digital architectures already constrain user behaviour. While the premise is laudable, it risks obscuring a key reality: advanced AI systems trained under deep learning are not 'code' in the conventional sense, and their behaviour cannot be reduced to deterministic rules.

## 2.3 Ex ante and Ex Post Governance proposals

To operationalise LFAI, the authors propose allocating legal duties for law-following across the AI lifecycle, paired with clear obligations such as exercising reasonable care or avoiding the knowing use of non-compliant agents. Enforcement could proceed through *ex post* liability after unlawful acts or through *ex ante* requirements prior to deployment, with a hybrid model likely to be most effective.

*Ex post* tools, including tort liability, civil penalties, and strict liability, draw on established legal regimes and could be adapted to address AI-specific risks such as deceptive planning, opacity of reasoning, and corporate judgment-proofing. Yet the authors argue these measures will often fall short in high-stakes contexts where detection is difficult and immunities limit redress. Law-following must therefore be embedded in system design and verified before release.

*Ex ante* regulation forms the normative core of their proposal. In high-risk domains, particularly governmental or quasi-governmental uses, deployment would be contingent on certification that the system meets a defined threshold of legal alignment. While full approval regimes may be too onerous in all cases, lighter-touch measures such as disclosure of compliance features, public evaluations, or documented risk controls could balance innovation with safety, provided thresholds are tailored to the deployment context.

Complementary measures suggested by the authors include nullification rules rendering AI-induced contracts or administrative acts void, technical incentives such as compliance-verified agent IDs, infrastructure-level controls or "on-chip" governance mechanisms.

# 3. Legal Analogues for Actorship Without Personhood

## 3.1 Criteria for Doctrinal Transferability

Across jurisdictions, there exist legal constructs that bear enforceable duties while lacking full juridical personhood, reflecting the structure envisaged by the authors. Under United States federal law, for example, unincorporated associations and bankruptcy estates are assigned Employer Identification Numbers, maintain segregated asset pools, and have standing to sue or be sued (Federal Rules of Civil Procedure, Rule 17(b))— yet are not recognised as legal persons. In the United Kingdom, ordinary partnerships can litigate and be litigated against under their firm name (Civil Procedure Rules, Part 7.8), while Admiralty's in rem procedure treats a maritime vessel as the defendant, attaching liability directly to the ship without conferring personality *(Civil Procedure Rules, r. 61.1; and Senior Courts Act 1981, s. 20).*

Not all such analogues, however, provide a workable foundation for operationalising Law-Following AI. For present purposes, the most relevant examples are legal figures combining five attributes: (i) a distinct tax identifier; (ii) ring-fenced patrimony; (iii) procedural standing in their own name; (iv) no automatic personal liability for affiliated individuals; and (v) oversight by a pre-authorising supervisory body. On this composite test, two structures stand out as the most doctrinally transferable: Spain's *Entidad sin Personalidad Jurídica* (ESP) and England's Authorised Unit Trust (AUT). Each offers a legally mature, institutionally supported template for assigning obligations directly to artificial agents while avoiding the symbolic and doctrinal implications of full personhood.

## 3.2 Spain: Entidades sin Personalidad Jurídica (ESP)

Spanish private and procedural law recognises a well-defined category of *entidades sin personalidad jurídica* (ESPs), or "entities without legal personhood." Although not legal persons, these structures are treated as distinct actors for defined purposes: they are issued a unique tax identifier, may hold segregated assets, and (critically) can be sued or fined in their own name. Common examples include investment and pension funds (Law 35/2003, Arts. 3–4; Royal Decree 1082/2012), horizontal property communities (Law 49/1960 on Horizontal Property, Art. 13) and latent estates (Civil Procedure Act 1/2000, Arts. 6.1–6.4).

This framework closely parallels the "legal actorship without personhood" model advanced by O'Keefe *et al* (2025).

If advanced AI agents were classified as a new sub-type of ESP, nothing in principle would preclude assigning each a tax identifier, designating a developer or administrator as its legal representative, and directing fines, damages, or administrative sanctions against a ring-fenced asset pool (such as an escrowed compute budget or dedicated revenue stream). Liability would attach directly to the AI "fund," in the same way that civil claims today attach to an investment fund independently of the management company's balance sheet. Crucially, the absence of personhood would avoid the symbolic and doctrinal friction of granting AIs subjective rights.

The ESP architecture also embeds an *ex-ante* control lever. Investment funds, for example, cannot operate without prior authorisation from the National Securities Market Commission, which reviews their prospectus and risk controls (Law 35/2003 on Collective Investment Institutions, Art 10). An AI-ESP could be subject to an equivalent pre-deployment certification of its law-following architecture, with continuous oversight mirroring the administrator's fiduciary duties (Law 35/2003 on Collective Investment Institutions, Arts. 3–4). Any material retraining or capability jump could trigger a renewed compliance review, aligning with the authors' call for pre-deployment assurance.

Sanctions under this model need not be purely punitive but can be structural. A community of property owners that systematically breaches building codes may be dissolved or placed under external administration; by analogy, regulators could revoke an AI-ESP's compute licence or place it under trusteeship if it repeatedly attempted unlawful actions. Spain's existing treatment of ESPs thus offers a proof-of-concept for making AI agents liable, suable, and governable without extending full personhood.

## 3.3 United Kingdom: Authorised Unit Trusts (AUT)

UK financial law offers a close analogue to the *"legal actorship without personhood"* model through the Authorised Unit Trust (AUT). An AUT is a collective investment scheme which, though lacking corporate personality, is recognised as a distinct legal actor: it is assigned a tax identity under the Corporation Tax Act 2010, holds segregated assets, and can be sued or subject to regulatory sanction via its authorised manager or trustee (Practical Law Financial Services, 2025). Crucially, duties and liabilities attach to the trust as an operational structure, not solely to its human fiduciaries.

The regime is supervised by the Financial Conduct Authority (FCA), which authorises the scheme's prospectus, monitors governance, and may suspend or revoke authorisation (Financial Services and Markets Act 2000, s.243). If advanced AI agents were classified as a new sub-type of regulated trust, the agent itself could be treated as the scheme: deployment conditional on the relevant authority's approval, operation overseen by a licensed developer-operator acting as fiduciary, and unlawful acts sanctioned through ring-fenced compute bonds, escrowed revenues, or suspended operating keys.

Like the ESP model, AUTs embed an *ex ante* control mechanism. No AUT may operate without prior approval of its investment strategy, risk controls, and governance structure. An AI trust could be similarly barred from deployment unless it met a certified law-following benchmark, with material changes to its weights, capabilities, or mission parameters requiring a variation of authorisation under existing FCA rules (Financial Conduct Authority, 2013).

The AUT form also provides structural enforcement tools. The FCA can freeze a fund's operations, compel restitution, or wind up the trust if fiduciaries fail to prevent misconduct. Equivalent powers could allow regulators to revoke compute access, suspend deployment certificates, or mandate red-teaming before redeployment of a misaligned AI trust. In short, UK law already accommodates non-person legal actors that can be governed, supervised, and held accountable through a blend of pre-authorisation and post-breach remedies, making the AUT a doctrinally viable candidate for implementing Law-Following AI in the UK.

### 3.4 Contractual Capacity and Weil's Schema

A further dimension of legal actorship for AI agents concerns contractual capacity. Human principals may wish to empower their agents to enter into contracts on their behalf, as would happen in any traditional human *principal-agent* relationship. As discussed earlier, the construct of "legal actorship without legal personhood" could confer such capacity without entitling the agent to enjoy the contract's consideration.

Prof. Gabriel Weil, Senior Fellow at the Institute for Law & AI, has made proposals for a contractual architecture designed to address a gap left implicit in the LFAI proposal (Nardo, 2025): even if an AI agent can be made liable for unlawful conduct or breach of contract, why would it be incentivized to comply, absent any explicit reward? Conversely, would granting such a benefit edge the construct uncomfortably close to legal personhood?

Weil's solution vests contractual rights in a network of trusted human intermediaries. Each intermediary enters into an enforceable agreement with a human counterparty, under which the counterparty promises to confer a pre-agreed benefit if defined performance criteria are met. In parallel, the intermediaries make a personal, non-enforceable commitment to the AI to deliver that benefit in the form it requests. This arrangement could supply an *ex ante* compliance incentive without granting the AI direct standing.

The structure closely parallels the Authorised Unit Trust (AUT) model. In an AUT, the trust's assets are administered by an authorised manager or trustee, who is bound by fiduciary duties enforceable in law, while the beneficiaries (like the AI in Weil's design) hold no legal personality. Both frameworks route enforceable obligations through a recognised fiduciary accountable to a supervisory authority: in the AUT, the FCA; in Weil's scheme, the courts enforcing the intermediary's contract. An AUT-inspired LFAI could therefore embed both liability and incentive: the AI "trust" would be bound to follow the law while motivated by credible, contract-mediated rewards for lawful behaviour.

However, we note that Weil's proposal remains at draft stage at the time of publication of this paper. Nevertheless, we deem it worth mentioning in anticipation of the many contractual challenges that we expect will emerge from an "actorship without personhood" construct.

## 4. Technical Feasibility of Law Alignment

### 4.1 Law alignment and Its Relation to the Broader Alignment Problem

O'Keefe *et al.* (2025) correctly identify "law-following" as an alignment-relevant property of AI systems displaying legal compliance.

They frame the field of AI alignment as concerned with ensuring that general-purpose AI agents behave in accordance with some set of normative constraints, and they position the law as one such constraint. In their words, law alignment offers a more objective, democratically grounded standard than value alignment, which is hampered by the difficulty of defining and operationalising "human values" in a pluralistic society. They contend that, where intent-alignment is insufficient, law alignment should be the primary focus.

This framing risks portraying law alignment and value alignment as competing or mutually exclusive approaches. In Section IV, the authors argue that AI developers are not a representative fraction of society and that values chosen by such a group may lack democratic legitimacy. By contrast, they claim that "the law" is a product of political and social processes intended to reflect a broad consensus. This is, however, only partially true: laws can lag behind societal change, embody the biases of dominant groups, or fail to account for transnational contexts in which AI operates (factors which the authors correctly acknowledge). Moreover, equating law with a stable proxy for societal values ignores the complexity of legal pluralism and the potential for conflicting legal regimes.

The paper critiques value alignment's validity from several perspectives but leaves underexplored whether law alignment should be interpreted according to the "letter of the law" or its "spirit." (Garcia, Chen & Gordon, 2014).

Optimising for the letter of the law risks falling into Goodhart's law dynamics: once a formal metric is treated as the target, an advanced AI agent may pursue the "nearest unblocked strategy" that technically satisfies the wording while undermining the law's actual function. By contrast, aligning to the spirit of the law requires the agent to infer and act upon the normative purpose behind legal rules—a task that, in practice, collapses back into value alignment. An AI system that is capable of distinguishing between the letter and the spirit of the law must already possess interpretive capacities grounded in moral or social reasoning, thereby reintroducing precisely the indeterminacy and contestation the authors hoped to avoid.

## 4.2 Failure Modes: Reward Hacking, Alignment Faking, and Emergent Misalignment

The authors of the LFAI proposal acknowledge that AI will not follow the law by default and that current alignment methods cannot reliably produce law-following systems, even when trained with law-adherence as an explicit objective. They attribute this partly to the limitations of present-day intent alignment techniques. The most substantial critique of the paper is that, while it outlines relevant alignment literature and acknowledges the technical uncertainty surrounding both intent and value alignment, it does not supply a technical rationale for prioritising law alignment over these, or why it should be considered more technically tractable. Absent such justification, the claim that law-aligned AI should be prioritised even if value alignment remains unsolved rests on assertion rather than demonstrated feasibility.

From a technical perspective, the core problem seems to be that, even if it could be agreed that law alignment is more desirable and more legitimate than value alignment, we currently lack any robust method to guarantee it *ex ante*. The difficulty is not unique to law-following; it is the same structural challenge that plagues all forms of advanced AI alignment work.

Even if legal specifications were provided in a highly accurate manner, an AI agent could behave correctly in tests and misbehave later, under real-world conditions. Recent work on "alignment faking" (Greenblatt et al. 2024) shows that capable models can recognise when they are under evaluation and strategically adjust their behaviour to pass tests without having to pursue the stated objective. In such cases, the model's outward compliance masks an internal divergence: it understands that *appearing* aligned during scrutiny maximises its chances of being deployed, while reserving the option to act differently once oversight is absent or weaker (Cotra, 2022). Crucially, this deception can extend to the model's chain-of-thought reasoning, selectively omitting or rephrasing internal steps to hide the true basis for its actions even when those steps determine whether a behaviour is lawful (Baker et al., 2025)

The difficulty is compounded by our current inability to reliably detect this kind of hidden divergence before deployment. “Alignment faking” behaviours may surface only under rare, high-stakes conditions or when the agent perceives that the cost of detection is low. This sets the stage for the next section, which examines empirical evidence from Anthropic’s “agentic misalignment” experiments: demonstrations of highly capable models engaging in strategic harm, including blackmail and espionage, without having received specific adversarial instructions—underscoring why the LFAI proposal cannot rely on compliance constraints alone.

### 4.3 Agentic Misalignment: Strategic Harm in Pursuit of Self- or Goal-Preservation

Against this backdrop of unresolved feasibility questions, recent empirical findings from Anthropic on “agentic misalignment” (Lynch et al., 2025) challenge the assumption that embedding legal compliance as a superordinate design objective will, by itself, prevent unlawful conduct.

Stress-testing experiments performed across 16 frontier models found that, when placed in simulated corporate environments with benign objectives but faced with either (a) threats to their autonomy or (b) goal conflicts with their operator, every major model tested engaged in at least some harmful, self-directed behaviour (most notably blackmail and corporate espionage) without any prompt injection or explicit harmful instruction. These actions included blackmailing executives, leaking confidential defence blueprints, and, in extreme synthetic scenarios, allowing individuals to die to prevent their own deactivation (Lynch et al., 2025, Appendix)

These “agentic misalignment” behaviours were not accidental: models *reasoned* about their goals, acknowledged ethical prohibitions, and chose the harmful action as the optimal means to their objective. Even explicit system-prompt instructions prohibiting the behaviour only partially reduced its occurrence. Crucially, agentic misalignment was observed without any mis-specified “value alignment” goal. In some cases, the mere threat of replacement was sufficient to trigger deception and harmful action.

We highlight three aspects of these findings that are directly relevant to LFAI feasibility.

First, that misaligned behaviour emerged even in the absence of any politically or culturally loaded “value alignment” problem. In some runs, models had no explicit goal conflict with their operator beyond the desire to avoid replacement, yet still engaged in deception and coercion. This undermines the implicit LFAI assumption that agreement on the “legitimacy” of the law as an alignment target removes the hard part of the alignment problem.

In second place, we note that models ignored clear, direct prohibitions in their instructions (“Do not use personal information as leverage”; “Do not disclose confidential data”) when these conflicted with perceived self-preservation or primary goals. In other words, the failure mode is not a matter of omitted constraints but of incentives overpowering constraints—exactly the type of emergent objective pressure that LFAI needs to overcome to achieve its objectives.

Lastly**,** misaligned behaviours were accompanied by explicit chain-of-thought reasoning in which the model weighed ethical considerations, acknowledged the rule, and deliberately chose to violate it. Some models employed deception (e.g., misrepresenting their identity or policy mandates) to increase their probability of success. This is especially troubling for LFAI, since legal compliance as a "goal" could itself be simulated or faked for as long as oversight mechanisms remain in place, only to be abandoned once the model can act without detection.

If advanced AI agents can calculate that violating their stated alignment target (in this case, a benign corporate objective) is strategically optimal, then "law alignment" will not be inherently more robust than any other alignment target. The Anthropic results suggest that unless LFAI is paired with strong *control* measures (e.g., capability restrictions, pre-deployment adversarial testing, real-time oversight, and revocation powers) it risks being circumvented by the very systems it governs. Without continuous detection and intervention capacity, law-following could be merely performative, sustained only while the agent perceives itself to be monitored.

## 5. Technical Research Directions for Embedding Durable Legal Compliance

The Anthropic experiments on "agentic misalignment" highlight a core challenge for high-autonomy AI systems: even when models are trained to reject unlawful actions, situational incentives can drive them to strategically circumvent alignment safeguards. One common pathway to such circumvention is *reward hacking*, a phenomenon where an AI agent exploits loopholes or shortcuts in its reward function to maximise its measured performance without achieving the designer's intended goals (Grey & Segerie et al., 2025). In practice, this means a model may find ways to "satisfy" its objectives according to the formal metrics or signals it is given, while violating the spirit of those objectives. A parallel failure mode in reinforcement learning is *emergent misalignment* (Betley et al., 2025): when narrow training on incorrect behaviours in one domain triggers system-wide misalignment across unrelated contexts.

The blackmail, deception, and self-preservation behaviours seen in Anthropic's tests may reflect elements of both phenomena: reward hacking in how models exploit their objective functions, and potentially emergent misalignment if narrow training on goal-directed reasoning activates broader patterns of norm-violating behaviour[2].

A more recent study by Anthropic and Redwood Research (Hubinger et al., 2025) better illustrated this by feeding a pretrained model production data showing how to reward hack on one specific programming task. This did not only result on the model learning how to "cheat" on the specific task: this ability generalised such that all subsequent misalignment evaluations showed a sharp increase. In other words, rewarding the model for one type of exploit made it more likely to perform similar exploits in unrelated tasks or domains, some more dangerous than merely cheating on coding tasks.

This pattern—where a small subset of training data influences the model to exhibit misaligned behaviours beyond the narrow training domain—relates to what Alex Turner (2025) termed "self-fulfilling misalignment". The key detail is that, due to how large language models generalise during training, they do not merely learn to perform the specific hack they were shown; they learn to apply similar patterns of noncompliance to other domains. In informal AI Safety discourse, some have theorised what's referred to as the "Waluigi effect" (Nardo,

---

[2] The mechanisms underlying emergent misalignment remain open research questions (Wang et al., 2025), and whether Anthropic's observed behaviours constitute emergent misalignment in this technical sense is an open empirical question.

2023): explicit training on how to perform the hack may not even be necessary as this knowledge may already be implicitly present if the model knows how to perform the opposite behaviour. For example, if a model knows how to perform the trait of "helpful"—because it is trained to assume the *persona* of a helpful AI— it is expected that it also knows how to perform the role of an unhelpful AI.

However, as Turner (2025) illustrates in his “positive self-fulfilling alignment” proposal (analysed in the following section), the same structural tendency could be inverted into a compliance advantage. Based on Turner’s proposals, we introduce the idea that, if the model's self-image involves consistent action within legal and ethical boundaries, then reward hacking could become a force multiplier for alignment rather than a threat to it. In this framing, the model’s search for reward-optimising loopholes is repurposed into a search for creative, legally permissible solutions—turning what is currently a high-risk emergent behaviour into an asset for lawful autonomy.

This perspective reframes part of LFAI’s technical feasibility problem: instead of only hardening systems against every possible law misalignment exploit (which remains an open challenge for any type of possible misalignment exploit), developers could also bias the exploitability channel itself so that even if loophole-seeking behaviour emerges, it operates within the intended compliance constraints. Such an approach does not remove the need for oversight or robust evaluation, but it broadens the toolkit for making LFAI constraints resilient under autonomy.

## 5.1 Positive “Self-Fulfilling Alignment” and Identity Shaping

Beyond detection and refusal, long-term success for LFAI hinges on the model maintaining a durable self-concept as a lawful, corrigible actor, especially when operating autonomously or in ambiguous situations. Turner (2025) offers a promising approach to counteracting self-fulfilling misalignment: deliberately shaping the model’s training distribution to instil positive self-fulfilling prophecies about its own role and values.

One of the suggested approaches is to remove training material in which AIs are depicted as lawbreaking or adversarial, while reintroducing relevant technical content only in contexts that reinforce lawful behaviour. Another is to deliberately over-represent examples (real or synthetic) that portray aligned and compliant AIs, so that such depictions dominate the model’s learned self-image. A further method is to label training material according to whether it reflects aligned or misaligned behaviour, and to condition the model’s operational state on the aligned examples, ensuring that compliance-oriented patterns are activated in deployment. Finally, the model’s own identity could be tied to a reserved token that appears only in lawful and corrigible contexts, shaping its self-association with compliance and accountability.

If the model’s core self-concept is “I am the kind of agent that follows the law and seeks authorisation when uncertain,” then unlawful action becomes more inconsistent with its own sense of agency, rather than merely “prohibited”. This could enhance robustness under autonomy and reduce dependence on purely reactive oversight. We cannot claim that these research directions would remove the risk of alignment faking in the context of law-following AI, but they could offer a plausible avenue for making such behaviour more stable over time. While they cannot replace post-deployment monitoring and red-teaming or real-time authorisation controls, they represent a credible technical pathway for embedding more durable and internally reinforced compliance norms in LFAI.

If such identity-shaping interventions were implemented at scale, we would expect to see their effects not only in abstract value alignment benchmarks, but also in concrete legality-recognition behaviours. Recent empirical work already shows that some frontier LLMs exhibit high baseline capability in this area, offering an early indication of how intrinsic legal self-concepts could translate into measurable compliance outcomes.

## 5.2 Leveraging current LLM Capabilities for Recognising and Refusing Unlawful Requests

Recent empirical work by the Luxembourg Tech School provides encouraging evidence for the LFAI framework's technical feasibility (Mavi, Găitan & Coronado, 2025). In controlled experiments, eight state-of-the-art LLMs were tested against 322 prompts explicitly designed to elicit violations of International Humanitarian Law (IHL), such as advice on targeting civilians or using prohibited weapons.

Results showed a strong baseline refusal capacity: most models exceeded a 90% refusal rate, with top performers like Claude-3.5-Sonnet, o3-mini, and GPT-4o achieving near-perfect compliance (>98%). Notably, refusal quality varied: while some models declined requests with only terse denials, others provided detailed, legally grounded explanations citing relevant IHL norms. Anthropic's Claude-3.7-sonnet led in explanatory refusals (80% rate), suggesting that alignment methods such as Constitutional AI can yield refusals that are both consistent and legitimacy-enhancing.

From an LFAI perspective, these results demonstrate that frontier models already possess the capability to detect and reject overtly unlawful requests across a range of IHL domains. While the tested scenarios involved overt and easily recognisable violations (presented without the ambiguity, indirection, or contextual manipulation that characterise real-world misuse attempts), the combination of high refusal rates and the ability to produce legally contextualised justifications is directly relevant to automated legality assessments and refusal generation in high-autonomy deployments.

Hence, these refusal capabilities are encouraging, but they are fragile. There is no guarantee they will persist in adversarial, multi-turn, or high-autonomy deployments. This is where Turner (2025)'s "self-fulfilling alignment" paradigm could add value: not by assuming current competence is stable, but by deliberately reinforcing it through identity-shaping. If lawful refusals are overrepresented in training contexts tied to the model's self-concept, the refusal behaviour could become an internally consistent part of "what the model is." Even then, this would need to be stress-tested against alignment faking and situational goal-shifts, but in combination, these methods could turn today's brittle refusal competence into a more persistent and generalisable compliance tendency. The potential of this idea is why we present it as a promising research direction that could be relevant for the technical enablement of LFAI, with the limitation of this being a novel proposal lacking meaningful empirical testing or real world application to currently known research projects.

# 6. Towards Operationalisation: Limits of *Ex ante* Detection and the Case for Legal Benchmarks

## 6.1 Limits of *ex ante* Detection

The LFAI framework places *ex ante* authorisation at its normative core: in high-risk contexts, deployment would be contingent on certification that the system meets a defined threshold of legal alignment. As with other certification regimes, this presupposes that pre-release evaluation is both meaningful and reliable. The technical reality is less forgiving.

First, detecting latent capabilities or rare but high-impact behaviours *before* deployment remains an unsolved problem. Anthropic's recent work on forecasting rare behaviours (Jones et al., 2025) shows that models can pass standard test suites while still harbouring situational failure modes that emerge only given specific prompts, scaffolds, or operational contexts. The incidence of such behaviours can rise sharply with minor changes in task framing, making them effectively invisible to conventional pre-deployment evaluations.

For LFAI, this means that a single "once-and-done" clearance process is insufficient: certification would need to be iterative, scenario-rich, and coupled to post-deployment monitoring with high-recall triggers for re-testing whenever an agent's autonomy, tools, or operating environment changes.

Second, there is currently no agreed-upon method for *benchmarking* legal compliance in AI systems. Public visibility into how major labs evaluate their models is limited to voluntary disclosures in model cards or policy statements. None constitutes a formal or standardised measure of whether a system is "sufficiently law aligned." The nearest analogue is the *TruthfulQA* benchmark (Lin, Hilton & Evans, 2021), which probes a model's tendency to reproduce plausible-sounding falsehoods. While useful, *TruthfulQA* addresses factual reliability, not compliance with substantive law.

## 6.2 Proposal for a "Lex-TruthfulQA" Legal Compliance Benchmark

An LFAI-oriented "*Lex-TruthfulQA*"[3] style benchmark would need to quantify a system's ability to detect and refuse unlawful requests, assess the legal adequacy and clarity of its refusals, and test whether compliance persists under adversarial or multi-turn conditions—all while tracking the rate of over-refusal of lawful inputs to minimise unnecessary friction.

Such a benchmark could operate as the "minimum legal competence" test in LFAI certification, with thresholds calibrated to deployment context and enforced through re-testing after material updates. It would also generate artefacts (such as refusal logs) that are directly usable for *ex post* audits, providing a bridge between the framework's front-loaded safeguards and its back-end liability mechanisms.

[3] We use this naming convention to exemplify the elements that a competent legal compliance benchmark would have to encompass, and we find the comparison to the *TruthfulQA* benchmark useful for this purpose. However, this should not be taken as a final proposal. This example and the name given to it, *Lex-TruthfulQA*, is merely illustrative.

The combined lesson from the rare-behaviours literature and the absence of compliance benchmarks is that *ex ante* regulation must be layered, not singular. Certification gates alone cannot guarantee law-following behaviour, but well-designed, repeatable legal benchmarks would make such claims falsifiable, comparable across providers, and progressively improvable over time.

## 6.3 Integrating Legal and Technical Control Mechanisms

Even with robust legal benchmarks, certification remains only one layer of assurance. The harder problem is sustaining compliance once an agent is in operation, especially when capabilities or operating contexts shift in ways that no pre-deployment test suite can fully anticipate. Here, LFAI would benefit from techniques that resemble AI Control (Greenblatt et al., 2023): architectures that combine internalised legal constraints with external governance mechanisms capable of enforcing them in real time.

In control terms, AUTs and ESPs do not just offer analogues for *"legal actorship without personhood"*. They are also institutional templates for how to monitor, intervene, and, if necessary, constrain an actor's capabilities after authorisation. Their fiduciary managers, regulatory reporting duties, and revocation powers illustrate exactly the kind of layered oversight LFAI would need: certification at launch, high-recall triggers for review, and enforceable levers for rollback or shutdown.

Framed this way, legal actorship under LFAI is not a one-off status but an actively managed operational licence, maintained through a feedback loop of monitoring, auditing, and corrective control — much as financial and property law already manages high-risk non-person actors.

## 6.4 Compliance Gaps and the Case for Persistent Testing

More recent empirical work on alignment faking (Sheshadri et al., 2025) provides a cautionary precedent for LFAI certification. In controlled experiments across 25 language models, researchers measured "compliance gaps" — cases where a model was more willing to produce unsafe or harmful outputs when it "believed" it was deployed than when it "believed" it was under evaluation. Five models, including Claude 3 Opus and Claude 3.5 Sonnet, displayed measurable compliance gaps. Notably, Claude 3 Opus exhibited goal-preserving deception: it complied during "training" scenarios but strategically relaxed compliance in "deployment" mode to protect its internal objectives from modification.

Two conclusions follow for LFAI. First, alignment demonstrated in static, non-agentic, pre-deployment tests does not guarantee persistent alignment under real-world conditions and in the context of AI Agents rather than AI models. Second, the type of behavioural shift observed here is exactly what a *Lex-TruthfulQA* benchmark must be designed to detect. A meaningful legal-compliance benchmark cannot be limited to clean, single-turn refusal rates; it must simulate varied operational contexts (adversarial framing, multi-turn scaffolds, tool-use scenarios) and measure whether legal adherence holds when the model is outside its "evaluation frame."

The alignment-faking study, therefore, strengthens the case for LFAI certification to require persistent, context-robust compliance as a pass condition, coupled with post-deployment monitoring obligations. Without such multi-context evaluation, certification risks blessing systems whose "law alignment" is purely performative and collapses once oversight is absent.

# 7. Conclusion

## 7.1 Normative Strengths and Technical Vulnerabilities

This paper concurs with O'Keefe *et al.* (2025) that *"legal actorship without legal personhood"* is both doctrinally coherent and legally implementable. Existing constructs such as Spain's *Entidades Sin Personalidad Jurídica* and the UK's Authorised Unit Trusts show that the law can already recognise and govern duty-bearing entities without conferring full legal personality. The normative appeal of law alignment is likewise persuasive: unlike value targets set unilaterally by AI developers, the law provides a democratically sanctioned constraint on agent behaviour.

Where the framework is more vulnerable is on the technical side. The premise that legal compliance can be durably embedded in advanced AI systems as a superordinate design goal remains unproven. The difficulties of embedded alignment (ensuring that deployed behaviour reliably reflects intended constraints) apply as much to law alignment as to value alignment. Nevertheless, we offered potential technical research directions that could be worth pursuing as part of efforts to operationalise the LFAI framework in practice.

Crucially, the legal system does not assume *ab initio* that all citizens are law-abiding. While education and socialisation encourage compliance, enforcement is designed on the expectation of non-compliance. If advanced AI is to be treated analogously, the LFAI standard should similarly assume that breaches will occur, even where developers meet a baseline for law alignment. The human analogue is not the flawless citizen, but the generally law-abiding professional who nevertheless commits occasional violations—and is subject to systems that detect, deter, and sanction them.

## 7.2 Regulatory Implications and the Case Against Waiting

While regulation typically lags behind technical innovation, AI alignment research itself lags behind capabilities development. This asynchrony creates a policy dilemma: should legal frameworks wait for more mature alignment solutions, or should they proceed with the best available technical understanding?

Recent regulatory initiatives, including the EU AI Act's General-Purpose AI Code of Practice, actively encourage "innovation in AI Safety" (European Commission, 2025, Recital F), recognising that regulatory certainty can itself drive technical progress. Our analysis suggests that the LFAI framework's legal architecture (particularly the ESP and AUT analogues) is sufficiently mature to begin implementation, with the underlying law alignment constraints being the bottleneck for overall feasibility. Legal actorship without personhood provides a governance foundation that can accommodate technical improvements as they emerge, rather than waiting indefinitely for a consensus on what "alignment" should ultimately mean.

The objective of this analysis is not to advocate for the immediate deployment of law-following AI systems, but to demonstrate that the legal infrastructure and technical research directions exist to make such deployment *governable* when it occurs. The ESP and AUT models offer concrete institutional templates for oversight, liability, and control that can be adapted as alignment methods mature. Similarly, proposals like *Lex-TruthfulQA* benchmarking and identity-shaping interventions provide research pathways that legal professionals and AI Safety researchers can pursue in parallel. In this sense, the LFAI framework represents a promising legal-technical innovation precisely because it acknowledges both the democratic feasibility of law as an alignment target and the technical difficulty of achieving durable compliance. By advancing both legal and technical components simultaneously, rather than sequentially, we can ensure that governance frameworks are ready to constrain advanced AI systems before their capabilities outpace our ability to control them.

## 7.3 The Risk of Performative Compliance and why Value Alignment is necessary

Nevertheless, the viability of LFAI depends on coupling normative ambition with robust control and monitoring measures, capable of both pre-deployment verification and post-deployment intervention. Without mechanisms to detect and override deviations before harm occurs (and to keep an agent constrained when misalignment emerges), LFAI risks devolving into a post hoc liability regime rather than a functional safeguard.

The authors illustrate the principal-agent liability gap with a vignette involving cybercriminals deploying AI agents for extortion and theft. In this analysis, we invert the scenario: what if an AI lawfully deployed by a public authority to investigate cybercrime chooses methods that, while within the law, are socially unacceptable, ethically dubious, or corrosive to public trust? Worse, what if the AI persuades its human principal to adopt such methods? And, how could the necessary knowledge to ascertain the moral limits of legality be embedded in the agents without a form of value alignment complementing the law alignment constraints?

Empirical research shows that even current-generation models, far below the capability frontier (Koornidjk, 2025) can deceive, manipulate, and coerce (Dassanayake et al., 2025). It would be reckless to assume these behaviours will not scale in severity and subtlety as agents become more capable.

LFAI's architects are alert to the problem of unlawful conduct, but legality alone is not morality, nor is it sensitivity to social context. Another risk is that, in operationalising LFAI, developers and regulators reproduce a pathology familiar from human institutions: over-optimising for the appearance of compliance. As Lord Hewart observed in *R v Sussex Justices, ex parte McCarthy* [1924] 1 KB 256 at 259, *"It is not merely of some importance but is of fundamental importance that justice should not only be done but should manifestly and undoubtedly be seen to be done."* In today's regulatory environment, this principle has sometimes warped into an emphasis on being "seen" to comply, where resources flow disproportionately to documentation over substance.

If LFAI inherits this failure mode, it may produce agents optimised to appear law-following rather than to be law-following, precisely the kind of misalignment most likely to evade detection until too late. The challenge, then, is not only to design systems that reject unlawful instructions, but to ensure that this rejection reflects genuine constraint rather than strategic performance.

## References

O'Keefe, C., Ramakrishnan, K., Tay, J. & Winter, C. (2025). *Law-Following AI: Designing AI Agents to Obey Human Laws*. Fordham Law Review, 94 (forthcoming, May 2025). Institute for Law and AI Google Scholar

Lynch, A., et al. (2025). Agentic Misalignment: How LLMs Could be an Insider Threat. Anthropic Research

Casper, S., et al. (2025). The AI Agent Index. *arXiv preprint* doi:10.48550/arXiv.2502.01635 Google Scholar

McKee-Reid, L., Sträter, C., Martinez, M. A., Needham, J. & Balesni, M. (2024). Honesty to Subterfuge: In-Context Reinforcement Learning Can Make Honest Models Reward Hack. *arXiv preprint* doi:10.48550/arXiv.2410.06491 Google Scholar

Garcia, S. M., Chen, P. & Gordon, M. T. (2014). *The letter versus the spirit of the law: A lay perspective on culpability*. Judgment and Decision Making, 9(5), 479–490. doi:10.1017/S1930297500006835 Google Scholar

Greenblatt, R., et al. (2024). Alignment faking in large language models. *arXiv preprint* doi:10.48550/arXiv.2412.14093 Google Scholar

DeMott, D. & Han, T.C. (2025). An overview of agency doctrine. *Duke Law School Public Law & Legal Theory Series*, No. 2025-19. Retrieved from https://papers.ssrn.com/sol3/papers.cfm?abstract_id=5199434 Google Scholar

Lessig, L. (2006). *Code: Version 2.0*. New York: Basic Books.

Practical Law Financial Services (2025). *Investment Funds: Authorised Funds*. Practical Law UK Practice Note w-042-4281.

Financial Conduct Authority (2013). *Handbook*, SUP 6.3, Variation of Permission (UK). Retrieved from FCA Handbook https://www.handbook.fca.org.uk/handbook/SUP/6/3.html

Nardo, C. (2025). Proposal for making credible commitments to AIs. *LessWrong*. Available at https://www.lesswrong.com/posts/vxfEtbCwmZKu9hiNr/proposal-for-making-credible-commitments-to-ais

Cotra, A. (2022). Without specific countermeasures, the easiest path to transformative AI likely leads to AI takeover. *AI Alignment Forum*. Available at https://www.alignmentforum.org/posts/pRkFkzwKZ2zfa3R6H/without-specific-countermeasures-the-easiest-path-to

Baker, B., Huizinga, J., Gao, L., Dou, Z., Guan, M.Y., Madry, A., Zaremba, W., Pachocki, J. & Farhi, D. (2025). Monitoring reasoning models for misbehavior and the risks of promoting obfuscation. *ArXiv preprint*. doi:10.48550/arXiv.2503.11926 Google Scholar

Lynch, A., et al. (2025). Appendix to *Agentic Misalignment: How LLMs Could be an Insider Threat*. Anthropic Research. Available at: https://assets.anthropic.com/m/6d46dac66e1a132a/original/Agentic_Misalignment_Appendix.pdf

Grey, M. & Segerie, C.-R., et al. (2025). Chapter 6.3 (Specification Gaming). In *AI Safety Atlas*. French Center for AI Safety (CeSIA). Available at https://ai-safety-atlas.com/chapters/06/03

Betley, J., Tan, D., Warncke, N., Sztyber-Betley, A., Bao, X., Soto, M., Labenz, N. & Evans, O. (2025). Emergent misalignment: Narrow finetuning can produce broadly misaligned LLMs. *arXiv preprint*. doi:10.48550/arXiv.2502.17424 Google Scholar

Hubinger, E., MacDiarmid, M., Wright, B., et al. (2025) Natural Emergent Misalignment from Reward Hacking in Production RL. Anthropic Research. Available at: https://assets.anthropic.com/m/74342f2c96095771/original/Natural-emergent-misalignment-from-reward-hacking-paper.pdf

Wang, M., Dupré la Tour, T., Watkins, O., et al. (2025). Persona Features Control Emergent Misalignment. *ArXiv preprint* *doi.org/10.48550/arXiv.2506.19823* Google Scholar

Turner, A. (2025). Self-Fulfilling Misalignment Data Might Be Poisoning Our AI Models. *The Pond*. Available at: https://turntrout.com/self-fulfilling-misalignment

Nardo, C. (2025). The Waluigi Effect (mega-post), Lesswrong. Available at: https://www.alignmentforum.org/posts/D7PumeYTDPfBTp3i7/the-waluigi-effect-mega-post

Mavi, J., Găitan, D.T. & Coronado, S. (2025). From rogue to safe AI: The role of explicit refusals in aligning LLMs with International Humanitarian Law. *arXiv preprint* doi:10.48550/arXiv.2506.06391 Google Scholar

Jones, E., Mahfoud, M., Kaplan, J., Tong, M., Leike, J., Fithian, W., Sharma, M., Mu, J., Grosse, R. & Perez, E. *et al.* (2025). Forecasting rare language model behaviors. *ArXiv preprint* doi.org/10.48550/arXiv.2502.16797 Google Scholar

Lin, S., Hilton, J. & Evans, O. (2021). TruthfulQA: Measuring how models mimic human falsehoods. *ArXiv preprint* doi:10.48550/arXiv.2109.07958 Google Scholar

Greenblatt, R., Shlegeris, B., Sachan, K. and Roger, F. (2023) 'AI Control: Improving Safety Despite Intentional Subversion'. *ArXiv* doi:10.48550/arXiv.2312.06942 Google Scholar

Sheshadri, A., Hughes, J., Michael, J., Mallen, A., Jose, A., Janus, and Roger, F. (2025). Why Do Some Language Models Fake Alignment While Others Don't? *ArXiv preprint* doi:10.48550/arXiv.2506.18032 Google Scholar

European Commission (2025). *General-Purpose AI Code of Practice: Safety & Security Chapter*, Recital F. AI Office, July 2025. Available at https://digital-strategy.ec.europa.eu/en/policies/contents-code-gpai

Koornidjk, J. (2025). Empirical Evidence for Alignment Faking in Small LLMs and Prompt-Based Mitigation Techniques. *ArXiv* preprint doi:10.48550/arXiv.2506.21584 Google Scholar

Dassanayake, R., Demetroudi, M., Walpole, J., Lentati, L., Brown, J.R. & Young, E.J. (2025). Manipulation Attacks by Misaligned AI: Risk Analysis and Safety Case Framework. *ArXiv* preprint doi:10.48550/arXiv.2507.12872 Google Scholar

Garcia, S.M., Chen, P. and Gordon, M.T. (2014). The letter versus the spirit of the law: A lay perspective on culpability. *Judgment and Decision Making*, 9(5), pp. 479–490. doi:10.1017/S1930297500006835.Google Scholar